\documentclass{article}

\usepackage[nonatbib, final]{neurips_2024}

\usepackage[utf8]{inputenc} 
\usepackage[T1]{fontenc}    
\usepackage{hyperref}       
\usepackage{url}            
\usepackage{booktabs}       
\usepackage{amsfonts}       
\usepackage{nicefrac}       
\usepackage{microtype}      
\usepackage{xcolor}         
\usepackage{adjustbox}
\usepackage{amsmath}
\usepackage{multirow}
\usepackage{color}
\usepackage{algorithm}
\usepackage{algorithmic}
\usepackage{setspace}
\usepackage{wrapfig}
\usepackage{enumitem}
\usepackage{pifont} 
\setlist{topsep=0pt, leftmargin=*}
\usepackage{graphicx}
\usepackage{caption}
\usepackage{subcaption}

\newtheorem{theorem}{Theorem}
\newtheorem{lemma}[theorem]{Lemma}
\newtheorem{proposition}[theorem]{Proposition}

\title{S-MolSearch: 3D Semi-supervised Contrastive Learning for Bioactive Molecule Search}

\author{%
Gengmo Zhou$^{1,2}$\thanks{\scriptsize Equal contribution.}, \ Zhen Wang$^2$\footnotemark[1], \ Feng Yu$^2$, Guolin Ke$^2$ ,
\textbf{Zhewei Wei$^1$}\thanks{\scriptsize Corresponding authors.}, \textbf{Zhifeng Gao$^2$\footnotemark[2]}  \\
$^1$Renmin University of China \quad $^2$DP Technology \\
\texttt{\{zgm2015, zhewei\}@ruc.edu.cn}, \texttt{\{wangz, yufeng, kegl, gaozf\}@dp.tech}
}

\begin{document}

\maketitle

\begin{abstract}

Virtual Screening is an essential technique in the early phases of drug discovery, aimed at identifying promising drug candidates from vast molecular libraries. 
Recently, ligand-based virtual screening has garnered significant attention due to its efficacy in conducting extensive database screenings without relying on specific protein-binding site information.
Obtaining binding affinity data for complexes is highly expensive, resulting in a limited amount of available data that covers a relatively small chemical space. Moreover, these datasets contain a significant amount of inconsistent noise. It is challenging to identify an inductive bias that consistently maintains the integrity of molecular activity during data augmentation. To tackle these challenges, we propose S-MolSearch, the first framework to our knowledge, that leverages molecular 3D information and affinity information in semi-supervised contrastive learning for ligand-based virtual screening. 
Drawing on the principles of inverse optimal transport, S-MolSearch efficiently processes both labeled and unlabeled data, training molecular structural encoders while generating soft labels for the unlabeled data.
This design allows S-MolSearch to adaptively utilize unlabeled data within the learning process.
Empirically, S-MolSearch demonstrates superior performance on widely-used benchmarks LIT-PCBA and DUD-E. It surpasses both structure-based and ligand-based virtual screening methods for AUROC, BEDROC and EF.

\end{abstract}

\section{Introduction}

Virtual Screening~\cite{klebe2006virtual, huang2018reverse, sydow2019advances, karaman2019computational} plays a crucial role in the early stages of drug discovery by identifying potential drug candidates from large molecular libraries.
Structure-Based Virtual Screening (SBVS)~\cite{meng2011molecular, lu2018computer,halgren2004glide, trott2010autodock}, a widely used virtual screening method, attempts to predict the best interaction between ligands against a protein target to form a protein-ligand complex. Recently, deep learning methods have also been explored. Methods trained on affinity labels ~\cite{ozturk2018deepdta, zheng2019onionnet, stepniewska2018development} conduct virtual screening by modeling binding affinities and ranking based on prediction. Additionally, a method~\cite{gao2024drugclip} uses similarities between embedding of pockets and molecules to search for active molecules. However, these SBVS methods cannot escape the dependency on the structure of protein targets, which is unavailable for challenging or novel targets, such as disordered proteins like c-Myc, limiting the applicability of SBVS. Besides, plenty of assays used in Virtual Screening~\cite{tran2020lit} are cell-based rather than target-specific, introducing noise into the active molecules since their activity is not entirely dependent on interaction with the protein target.

To remedy this, Ligand-Based Virtual Screening (LBVS)~\cite{hawkins2007comparison, sastry2011rapid, taminau2008pharao, yan2013enhancing, roy2014ligsift} searches similar molecules via known bioactive molecules and does not depend on the structure of protein targets, which has attracted increasing attention. Computational LBVS methods~\cite{liu2011shafts, lu2011shafts, roy2014ligsift} rigidly employ structural similarity to search for molecules, often using atom-centered, smooth Gaussian overlays to assess molecular similarity. Searching for bioactive molecules needs to consider both structure and electronic similarity. Some methods~\cite{hawkins2007comparison} also require charge comparison, which is expensive and time-consuming. These methods struggle with inefficiency when handling large databases. Moreover, since only structural or charge information is considered while affinity is ignored, even if molecules with similar structures or charges are identified, they may still exhibit poor affinity in practical applications due to activity cliffs~\cite{dimova2016advances}. 

A natural question arises: how can we enhance LBVS by collecting molecule similarity data from large-scale unlabeled molecules? To search both similar and bioactive molecules, we can choose those bioactive molecules that bind to the same protein. Meanwhile, labeled molecule-protein binding data is limited due to the expensive affinity acquisition. Also various standards~\cite{mysinger2012directory, tran2020lit} across different datasets, introducing substantial noise. It is difficult to cover the searching chemical space with affinity data alone. A feasible solution is to leverage similarity from finite affinity data to broader chemical space.

Inspired by the success of contrastive learning~\cite{chen2020simple, he2020momentum,radford2021learning, grill2020bootstrap}, which can extract informative representations from large-scale data, we explore its feasibility for molecule data. However, those data augmentation techniques in vision or language cannot be directly applied to molecules due to their inherent 3D structure. This limitation stems from a lack of inductive bias to guarantee augmentation maintains the integrity of molecular activity.

To address the challenges, we propose S-MolSearch, a novel semi-supervised contrastive learning framework based on inverse optimal transport (IOT)~\cite{stuart2020inverse}. S-MolSearch directly uses 3D molecular structures to capture structure similarity.
It consists of two main components: one encoder \(f_{\theta}\) for labeled dataset and another encoder \(g_{\psi}\) for the full dataset, which includes both labeled and unlabeled data. Both encoders are trained simultaneously to effectively leverage the two types of data. We organize a dataset of labeled molecule-protein binding data from ChEMBL~\cite{gaulton2012chembl}. Molecules are assigned to different targets based on binding affinity, with the active molecules corresponding to each target forming clusters. We sample from these clusters, treating molecules from the same cluster as positive samples and those from different clusters as negative samples to train encoder \(f_{\theta}\). This approach incorporates affinity information into training and avoids the limitation caused by relying solely on structural similarity. For the update of encoder \(g_{\psi}\), we assume that the similarity measurements obtained by encoder \(f_{\theta}\) can be generalized to unlabeled data. We input the same data from the full dataset into these two encoders separately and use optimal transport to obtain soft labels from \(f_{\theta}\). The encoder \(g_{\psi}\) is then trained using these soft labels.
This integration enables S-MolSearch to effectively utilize unlabeled data, ensuring that the model learns from both affinity-labeled samples and broader structural similarities across the molecular dataset. 

Empirically, S-MolSearch demonstrates superior performance on widely-used benchmarks LIT-PCBA~\cite{tran2020lit} and DUD-E~\cite{mysinger2012directory}. It consistently achieves state-of-the-art results, surpassing both SBVS methods and LBVS methods on AUROC, BEDROC and EF.

Notably, S-MolSearch trained with a 0.9 similarity threshold significantly outperforms existing methods, achieving more than a 49\% improvement on BEDROC and over a 30\% improvement on EF compared to the best baseline on DUD-E.
These results provide strong empirical support for the effectiveness of S-MolSearch, confirming its advanced capability for virtual screening.

Our main contributions are summarized as follows:
\begin{itemize}
    \item We introduce S-MolSearch, which is the first time integrates both molecular 3D structures and affinity information into molecule search.
    \item Built upon inverse optimal transport, we develop a semi-supervised contrastive learning framework, which induces S-MolSearch. By combining limited labeled data with extensive unlabeled data, S-MolSearch can learn more informative representations and explore the chemical space more effectively.
    \item S-MolSearch is evaluated on widely-used benchmarks LIT-PCBA and DUD-E, surpassing both SBVS and LBVS methods to achieve state-of-the-art results.
\end{itemize}

\section{Related work}

\subsection{Virtual Screening}

Virtual screening can be broadly divided into two main categories: Structure-Based Virtual Screening (SBVS) and Ligand-Based Virtual Screening (LBVS). SBVS~\cite{meng2011molecular, lu2018computer,halgren2004glide,trott2010autodock} heavily relies on the structure of protein targets and typically employs molecular docking. Recently, many deep learning methods~\cite{ragoza2017protein, zheng2019onionnet, stepniewska2018development, gao2024drugclip} have also emerged.
In contrast, LBVS~\cite{hawkins2007comparison, sastry2011rapid, taminau2008pharao, yan2013enhancing, roy2014ligsift} uses known active ligands as seeds to identify potential ligands. 
Molecule search is a major LBVS approach, typically divided into two categories: 2D similarity search and 3D similarity search. 2D molecule search methods~\cite{willett2006similarity, yan2012gsa} use molecular fingerprints to search for similar molecules, while 3D molecular search methods~\cite{sastry2011rapid, taminau2008pharao, roy2014ligsift} depend on shape overlap.

\subsection{Optimal Transport and inverse optimal transport}

Optimal Transport (OT) is a mathematical problem that aims to determine the most efficient way to redistribute one initial distribution (known as the source distribution) into another distribution (known as the target distribution) while minimizing a defined transportation cost.  
To handle computational complexities, OT often incorporates regularization~\cite{wilson1969use}, leading to a softened optimization problem.
The regularized OT objective is a convex function, thereby ensuring a unique solution that can be efficiently solved using iterative methods~\cite{sinkhorn1967diagonal, blondel2018smooth}.

Inverse Optimal Transport (IOT) seeks to determine the cost matrix that explains an observed optimal transport. ~\cite{stuart2020inverse} introduces a method to infer unknown costs. ~\cite{chiu2022discrete} explores the mathematical theory behind IOT. In many IOT studies~\cite{dupuy2016estimating,li2019learning}, optimization is directly performed over the cost matrix, typically focusing on learnable distances between samples rather than on the sample features.

\subsection{Semi-supervised learning}

Semi-supervised learning~\cite{learning2006semi, berthelot2019mixmatch, xie2020self, assran2021semi} is typically used in scenarios where labeled data is limited but unlabeled data is abundant. Pseudo-labeling~\cite{lee2013pseudo} is a classic technique of semi-supervised learning.~\cite{laine2016temporal} introduce a self-ensembling method that generates pseudo-labels by forming a consensus prediction using the outputs of the network under different regularization and augmentation conditions. UPS~\cite{rizve2020defense} proposes an uncertainty-aware pseudo-label selection framework that improves pseudo-labeling accuracy by reducing the amount of noise in the training process. UST~\cite{mukherjee2020uncertainty} employs a teacher-student training paradigm. The teacher model is responsible for selecting and generating pseudo-labels, while the student model learns from the labeled set augmented with these pseudo-labels. Recent work~\cite{mo2024s} utilizes additional unpaired images to construct caption-level and keyword-level pseudo-labels, enhancing training.
\section{Method}

\subsection{Overview}
Molecular similarity search is a type of ligand-based virtual screening whose purpose is to perform a rapid search and filtering of similar molecules in a molecular database based on a query molecule provided by the user. We model the task as a dense retrieval problem, using a well-trained encoder to extract embedding representations of molecules and rank them by their cosine similarity to a query molecule, thereby identifying the top $k$ most similar candidates.

\begin{figure}[t]
    \centering
    \includegraphics[width=\textwidth]{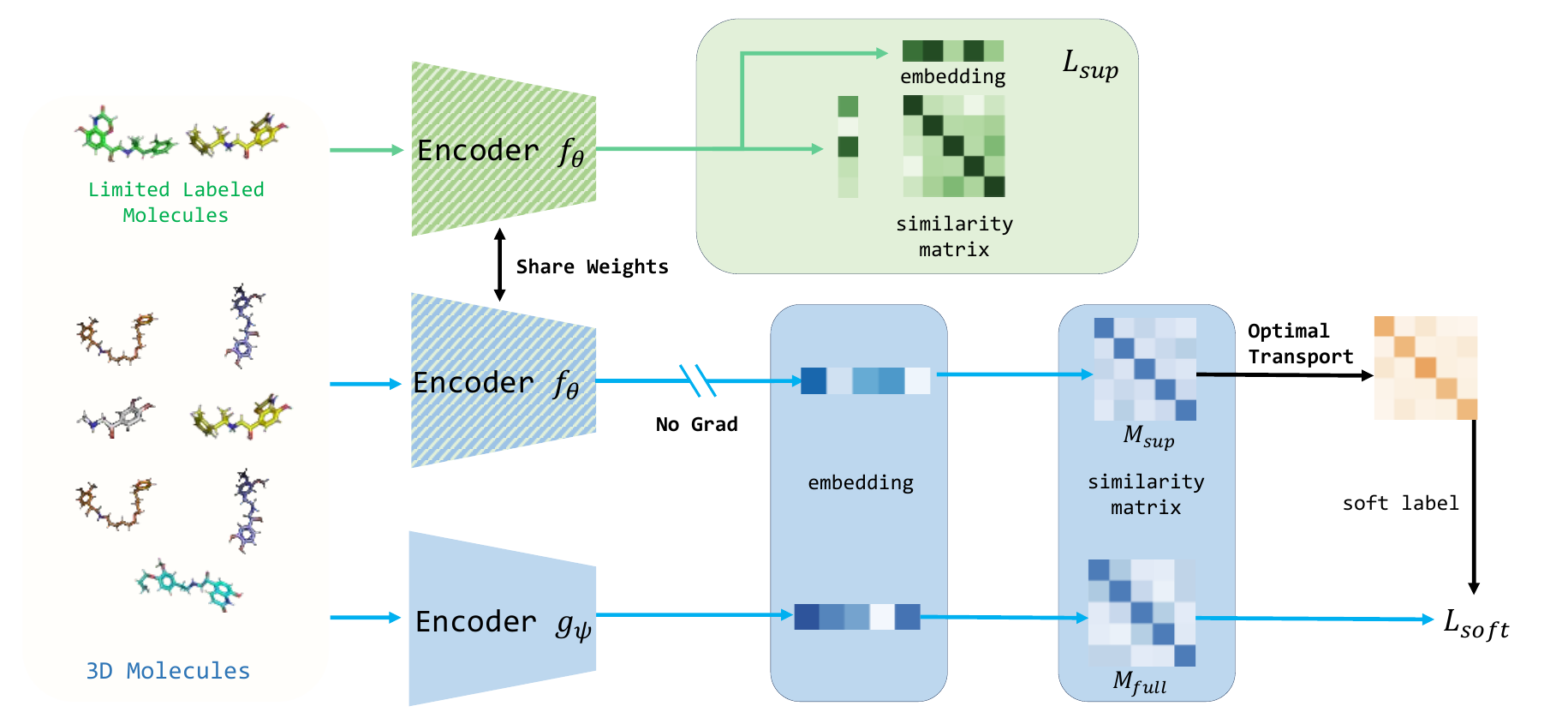}
    \caption{Overview of S-MolSearch Framework}
    \label{fig:overview}
\end{figure}

Building upon the principles of inverse optimal transport (IOT), we have developed the S-Molsearch method. As shown in Figure \ref{fig:overview}, S-Molsearch uses a molecular structure encoder \(f_{\theta}\) for labeled dataset \(D_{sup}\) and another encoder \(g_{\psi}\) for the full dataset \(D_{full}\), encompassing both labeled and unlabeled dataset. \(f_{\theta}\) utilizes contrastive learning to learn from labeled dataset. \(g_{\psi}\) optimizes its parameters using the soft labels produced by \(f_{\theta}\), which have been processed through smooth optimal transport. The two encoders are initialized with a molecular pretraining backbone Uni-Mol~\cite{zhou2023unimol}. Both encoders are trained simultaneously under the guidance of a unified loss function \( \mathcal{L}_{total}\). The encoder \(g_{\psi}\), trained on full dataset, is used for inference.

\graphicspath{{figures/}}

In the following sections, we will provide more details about the core components and training strategies that underpin S-Molsearch. In section 3.2, we explain the pretraining backbone of the molecular structural encoder, Uni-Mol. In sections 3.3 and 3.4, we will sequentially examine the training strategies for both \(f_{\theta}\) and \(g_{\psi}\). In section 3.5, we will discuss the regularization techniques employed to enhance model generalizability and stability. In section 3.6, we will provide an analysis of S-Molsearch from the perspective of IOT, offering insights into its methodological strengths.

\subsection{Pretraining Backbone of Molecular Encoder}

To effectively encode the structural information of molecules, we choose Uni-Mol as the backbone of molecular encoder for S-Molsearch. Uni-Mol is a molecular pretraining model specifically designed to adeptly process molecular 3D conformation data. It has achieved state-of-the-art performance across a range of downstream tasks. The UniMol model utilizes a self-attention mechanism that incorporates distance bias to integrate information about atoms and their spatial relationships, thereby generating a structural representation of molecules. The molecular embedding is created by using the embedding of the CLS token, and the embedding vector is normalized using the Euclidean norm.

\subsection{Training Strategy of Encoder on Labeled Dataset}

In this part, we employ molecular-protein binding data sourced from the ChEMBL. 
Molecules are assigned to different targets based on binding affinity. The active molecules corresponding to each target form clusters. We sample from these clusters to obtain data for contrastive learning. Molecules from the same cluster are considered positive pairs, while molecules from different clusters are considered negative pairs.
We employ InfoNCE loss for encoder \(f_{\theta}\) on labeled dataset, as shown in Equation \ref{eq infonce}:
\vspace{-2.5pt}
 \begin{equation}
 \label{eq infonce}
 \mathcal{L}_{sup} = -\sum_{i=1}^N \log \frac{\exp(\text{sim}(x_i, y_i) / \tau)}{\sum_{j=1}^N \exp(\text{sim}(x_i, y_j) / \tau)} 
 \end{equation}

where \( x_i \) and \( y_i \) are the embeddings of positive molecule pairs, while \( x_i \) and \( y_j \) form negative molecule embedding pairs within the batch when \( j \neq i \). \( \text{sim}(x, y)\) denotes the similarity score between embeddings, typically computed as the inner product of the normalized vectors \(xy^{T}\). \( \tau \) is the temperature parameter that scales the similarity scores. By utilizing the InfoNCE loss, \(f_{\theta}\) pulls the embeddings of positive samples close while enforcing them away from the negative samples in the embedding space.

\subsection{Training Strategy of Encoder on Full Dataset}

To harness large-scale unsupervised data effectively, S-MolSearch utilizes the \(f_{\theta}\) to guide the learning of another encoder on full data \(g_{\psi}\). 
Specifically, we employ \(f_{\theta}\) to preprocess the data for the \(g_{\psi}\), ensuring that \(f_{\theta}\) remains detached from the backpropagation process at this stage.
The embeddings obtained from \(f_{\theta}\) are subsequently utilized for computing a similarity matrix \(M_{sup} \in R^{N \times N}\), where \(M_{sup}(i,j) = x_{sup, i}x_{sup, j}^{T}\) The task of generating soft labels for \(g_{\psi}\) based on this similarity matrix is presented as a smooth and sparse optimal transport (OT) problem:
\vspace{-2.5pt}
\begin{equation}
\begin{aligned}
\min_{\Gamma \in U(p, q)} &\; \langle \Gamma, C \rangle + \frac{\lambda}{2} ||
\Gamma||^{2}  \\
 \text{subject to}& \quad U(p, q) = \{\Gamma \in R_{+}^{N \times N} | \Gamma \mathbf{1}_N = p, \Gamma^\top \mathbf{1}_N = q \} 
\end{aligned}
\end{equation}
Where \( \Gamma \) denotes the transportation plan matrix between the embeddings, \( C \in R_{+}^{N \times N} \) is the cost matrix derived from the cosine similarities between embeddings: \(C_{i,j} = c - x_{sup, i}x_{sup, j}^{T}\), \(\langle \Gamma , C \rangle\) denotes Frobenius inner product of \(\Gamma\) and \(C\), and \( p \) and \( q \) are the source and sink distributions, respectively. In this context, \(c = 1\) , \(p = 1_{N}\) and \(q = 1_{N}\). \(1_{N}\)
  is an \(N\)-dimensional vector of all ones. By introducing the OT formulation, we guarantee that signals from the supervised model \(f_{\theta}\) are more effectively transferred to the unsupervised model \(g_{\psi}\), while handling high label uncertainty in the supervised model \(f_{\theta}\) with appropriate regularization. The OT could be effectively addressed by the POT\cite{flamary2021pot}. We define \( \Gamma_{i,j} \) as pseudo-labels for the similarity matrix \(M_{full}\) of \(g_{\psi}\), where \(M_{full}\) is created by embedding of \(g_{\psi}\) \(M_{full}(i,j) =  x_{full, i}x_{full, j}^{T}\).

The cross-entropy loss \( H \) is employed to optimize \(g_{\psi}\), using the pseudo-labels provided:
\begin{equation}
\mathcal{L}_{soft} =  H(\Gamma, M_{full}) 
\end{equation}

\subsection{Regularization techniques}

In order to promote uniformity of embedding space, we apply KoLeo regularizer\cite{sablayrolles2018spreading, oquab2023dinov2} to the embeddings of the semi-supervised encoder. KoLeo regularizer is defined as:

\begin{equation}
\label{reg smol}
 \mathcal{L}_{reg} = -\frac{1}{n} \sum_{i=1}^n \log(\rho_{n,i}) 
\end{equation}
Here, \( \rho_{n,i} \) represents the minimum distance between the \( i \)-th sample and all other samples, which serves as a proxy for local density. This loss function has a geometric interpretation that effectively pushes closer points apart, ensuring diminishing returns as distances increase, thereby encouraging a uniformly dispersed embedding space.

\subsection{Framework for S-Molsearch Induced by Inverse Optimal Transport}
By integrating the losses and regularization terms from sections 3.3, 3.4, and 3.5, we have derived the overall loss function \( \mathcal{L}_{total}\) for the S-MolSearch model:
\begin{equation}
\mathcal{L}_{total} = \mathcal{L}_{sup} + \mathcal{L}_{soft}  +  \mu\mathcal{L}_{reg}
\end{equation}
where \(\mu\) is 0.1 in our setting. Building on the relationship between contrastive learning and IOT established in \cite{shi2023understanding}, we extend this relationship to a semi-supervised contrastive learning in proposition \ref{IOT-SCL-Framework}.

\begin{proposition}
\label{IOT-SCL-Framework}

Given encoder \(f_{\theta}\) for labeled dataset \(X_{sup}\) and \(g_{\psi}\) for full dataset \(X_{full}\), \(x_{sup}\) represents the embeddings of labeled data from \(f_{\theta}\) , while \(x_{full}\) represents the embeddings of the full dataset from \(g_{\psi}\). Semi-supervised contrastive learning is then formulated using IOT as follows:

\begin{equation}
\label{eq framework}
\begin{aligned}
& \min_{\theta, \psi} \left( KL(\Gamma^{g} || \Gamma^{\theta}) + KL(\hat{\Gamma}^{\theta} || \Gamma^{\psi}) + \mu Reg_{1}(\Gamma^{\psi}) + \nu Reg_{2}(\Gamma^{\theta})) \right) \\
& \text{subject to} \quad 
\begin{aligned}[t]
    & \Gamma^{\theta} = \arg\min_{\Gamma \in U(a), \, a = \frac{\mathbf{1}}{N}} \left( \langle C^{\theta}, \Gamma \rangle - \tau H(\Gamma) \right), \\
    & \Gamma^{\psi} = \arg\min_{\Gamma \in U(a), \, a = \frac{\mathbf{1}}{N}} \left( \langle C^{\psi}, \Gamma \rangle - \tau H(\Gamma) \right), \\
    & \hat{\Gamma}^{\theta} = \mathcal{T}(f_{\theta}^{fixed}, g_{\psi}^{fixed}, X_{label}, X_{full})
\end{aligned}
\end{aligned}
\end{equation}

where \( KL(X||Y) = \sum\limits_{ij}x_{ij}log\frac{x_{ij}}{y_{ij}} - x_{ij} + y_{ij}\) represents the Kullback-Leibler divergence, and \(H(\Gamma) = -\sum\limits_{i,j} \Gamma_{ij} (\log(\Gamma_{ij})-1)\) represents entropic regularization. 
\( \Gamma^{\theta}, \Gamma^{\psi}, \Gamma^{g} \in R_{+}^{N \times N}\),
\(\Gamma^{g}_{ij} = \frac{\delta_{ij}}{N}\) represents the ground truth based on labeled data,
\(\delta_{ij}\) denotes the Kronecker delta function.
\(C^{\theta}, C^{\psi} \in R_{+}^{N \times N}\)  are cost matrix of \(f_{\theta}\), \(g_{\psi}\) and \(C^{\theta}(i,j) = c -x_{sup,i}x_{sup,j}^{T}\), \(C^{\psi}(i,j) = c -x_{full,i}x_{full,j}^{T}\).  
\(\mathcal{T}\) generally refers to a technique for transferring supervised information to unsupervised data. 
\(Reg_{1}, Reg_{2}\) denotes regularization term. 
\end{proposition}

The proof is provided in the Appendix \ref{SCL-F}. 
In proposition \ref{IOT-SCL-Framework}, we model the contrastive learning problem on the labeled dataset and full dataset as two optimal transport problems. Additionally, we use \(\mathcal{T}\) to transfer knowledge from the labeled dataset to the unlabeled data, with the method of transfer depending on prior assumptions about the dataset and certain bias structures. For instance, If we initially optimize \(f_{\theta}\) on a large-scale labeled dataset to obtain \(f_{\theta}^{*}\), then generate \(\hat{\Gamma}^{\theta}\) as \(\hat{\Gamma}^{\theta}(i,j) = e_{full,i}e_{full,j}^{T}\), where \(e_{full,i} = f_{\theta}^{*}(x_{i}), x_{i} \in X_{full}\), we can develop a model that leverages knowledge distillation for contrastive learning. In the context of molecular search tasks, we employ smooth optimal transport for the modeling of \(\hat{\Gamma}^{\theta}\), leading to the development of S-MolSearch as follows:

\begin{proposition}
\label{IOT-SMOL}
Assuming the conditions outlined in Proposition \ref{IOT-SCL-Framework} are satisfied, the optimal parameters $\theta^{*}$ and $\psi^{*}$ of S-MolSearch can be regarded as the solution to the following IOT problem:

\begin{equation}
\begin{aligned}
& \min_{\theta, \psi} \left( KL(\Gamma^{g} \| \Gamma^{\theta}) + KL(\hat{\Gamma}^{\theta} \| \Gamma^{\psi}) + \mu Reg_{1}(\Gamma^{\psi}) \right) \\
& \text{subject to} \quad 
\begin{aligned}[t]
    & \Gamma^{\theta} = \arg\min_{\Gamma \in U(a), \, a = \frac{\mathbf{1}}{N}} \left( \langle C^{\theta}, \Gamma \rangle - \tau H(\Gamma) \right), \\
    & \Gamma^{\psi} = \arg\min_{\Gamma \in U(a), \, a = \frac{\mathbf{1}}{N}} \left( \langle C^{\psi}, \Gamma \rangle - \tau H(\Gamma) \right), \\
    & \hat{\Gamma}^{\theta} = \arg\min_{\Gamma \in U(a,b), \, a = \mathbf{1}_{N}, \, b = \mathbf{1}_{N}} \left( \langle C^{\theta^{\text{fixed}}}, \Gamma \rangle + \frac{\lambda}{2} \|\Gamma\|^{2} \right)
\end{aligned}
\end{aligned}
\end{equation}

where \(C^{\theta^{fixed}} \in R_{+}^{N \times N}\) and \(C^{\theta^{fixed}}(i,j) = c -x_{full,i}^{fixed}(x_{full,j}^{fixed})^{T}\). The \(x_{full}^{fixed}\) represents the embeddings of the same data in \(X_{full}\) obtained from the supervised encoder \(f_{\theta}\), where \(f_{\theta}\) is detached.
\end{proposition}
The proof is located in the Appendix \ref{SMOL}. We compute the KL divergence to guide the optimization of \(f_{\theta}\) and \(g_{\psi}\), where the regularization term simplification only affects the full data. Moreover, we set the marginal values of \(U(a,b)\) to an all-ones vector. In this way, we find that the knowledge in \(f_{\theta}\) transfers effectively to \(g_{\psi}\), thereby achieving excellent performance on molecule search task.

\section{Experiments}

\subsection{Training Data}
The labeled data comes from ChEMBL~\cite{gaulton2012chembl}, an open-access database containing extensive information on bioactive compounds with drug-like properties. 
We prepare nearly 600,000 protein-molecule pairs, encompassing about 4,200 protein targets and 300,000 molecules. The details of data curation can be found in appendix~\ref{app:chembl_data}. To prevent information leakage, the data is filtered based on protein sequence similarity. Specifically, the amino acid sequences of all protein targets in the benchmarks DUD-E and LIT-PCBA are extracted. Then, we use MMseqs~\cite{hauser2016mmseqs} tool with similarity thresholds of 0.4 and 0.9 to filter out proteins in ChEMBL. Using a 0.9 threshold helps filter out identical and highly similar targets, while the stricter 0.4 threshold filters out nearly all similar targets. After filtering, 3,369 proteins and 327,917 protein-ligand pairs remain for the 0.4 threshold, while 4,102 proteins and 529,856 pairs remain for the 0.9 threshold. We sample 1 million pairs from the filtered data as labeled data respectively.

The unlabeled data, consistent with what is used by Uni-Mol, comes from a series of public databases, totaling about 19 million entries. Additionally, we incorporate the small molecule data from ChEMBL into this collection, thereby obtaining the full dataset.

\subsection{Benchmarks}
We choose the widely used virtual screening benchmarks DUD-E~\cite{mysinger2012directory} and LIT-PCBA~\cite{tran2020lit} to evaluate the performance of S-MolSearch.
DUD-E is designed to help benchmark virtual screening programs by providing challenging decoys. It includes 102 protein targets along with 22,886 active ligands, each accompanied by 50 decoys with similar physico-chemical properties.
LIT-PCBA is designed for virtual screening and machine learning, aiming to address the chemical biases present in other benchmarks such as DUD-E. It consists of 15 targets, with 7,844 confirmed active compounds and 407,381 inactive compounds.

\subsection{Baselines}

We choose a range of LBVS and SBVS methods as comparative baselines for a thorough evaluation.
ROCS~\cite{hawkins2007comparison}, Phase Shape~\cite{sastry2011rapid}, LIGSIFT~\cite{roy2014ligsift}, and SHAFTS~\cite{liu2011shafts, lu2011shafts} are LBVS methods that evaluate similarity by calculating the overlap of molecular 3D shapes. Other methods are SBVS methods. Among them, DeepDTA~\cite{ozturk2018deepdta}, OnionNet~\cite{zheng2019onionnet}, Pafnucy~\cite{stepniewska2018development}, BigBind~\cite{brocidiacono2023bigbind}, and Planet~\cite{zhang2023planet} are trained on binding affinity labels. Glide~\cite{halgren2004glide}, Vina~\cite{trott2010autodock}, and Surflex~\cite{spitzer2012surflex} are molecular docking software. Gnina~\cite{mcnutt2021gnina} is a deep learning based molecular docking method. DrugClip~\cite{gao2024drugclip} utilizes the similarity between targets and molecules to find active compounds.

\subsection{Results}

\subsubsection{Main Results}

\begin{table}[ht]
  \centering
    \caption{Performance on DUD-E in zero-shot setting. The best results are \textbf{bolded} and the second-best results are \underline{underlined}.}
    \small
    \label{dude}
    \centering
    \small
    \begin{tabular}{l|l|l|lll}
      \toprule
      Method & AUROC (\%) & BEDROC (\%)  & EF 0.5\% & EF 1\% & EF 5\% \\
      \midrule
      ROCS & 75.20 & - & - & 23.79 & 6.89 \\
      Phase Shape & 76.70 & - & - & 30.33 & 9.01 \\
      LIGSIFT & 78.40 & - & - & 25.89 & 8.01 \\
      SHAFTS & 78.20 & - & - & 32.49 & 9.67 \\
      \midrule
      Glide-SP & 76.70 & 40.70 & 19.39 & 16.18 & 7.23 \\
      Vina &71.60  & - & 9.13 & 7.32 & 4.44 \\
      Pafnucy & 63.11 & 16.50 & 4.24 & 3.86 & 3.76  \\
      OnionNet & 59.71 & 8.62 & 2.84 & 8.83 & 5.40  \\
      Planet & 71.60 & - & 10.23  & 8.83  & 5.40 \\
      DrugCLIP & 80.93 & 50.52 & 38.07  & 31.89 & 10.66  \\
      \midrule
      S-MolSearch\textsubscript{{0.4}}& \underline{84.61} & \underline{54.22}  & \underline{40.85} & \underline{34.60} & \underline{11.44}  \\
      S-MolSearch\textsubscript{{0.9}}& \textbf{92.56} &  \textbf{75.37} & \textbf{51.50} & \textbf{47.94} & \textbf{15.82}  \\
      \bottomrule
    \end{tabular}
\end{table}

\begin{table}[ht]

    \caption{Performance on LIT-PCBA in zero-shot setting.}
    \label{pcba}
    \centering
    \small
    \begin{tabular}{l|l|l|lll}
      \toprule
      Method & AUROC (\%) & BEDROC (\%) & EF 0.5\% & EF 1\% & EF 5\% \\
      \midrule
      ROCS & 52.41 & - & - & 2.48 & -  \\
      Phase Shape & 52.24 & - & - & 2.98 & -  \\
      LIGSIFT & 54.94 & - & - & 2.39 & -  \\
      SHAFTS & 54.53 & - & - & 2.79 & -  \\
      \midrule
      Surflex & 51.47 & - & - & 2.50 & - \\
      Glide-SP & 53.15 & 4.00 & 3.17 & 3.41 & 2.01 \\
      Planet & 57.31 & - & 4.64 & 3.87 & 2.43 \\
      Gnina & \underline{60.93} & 5.40 & - & 4.63 & - \\
      DeepDTA & 56.27  & 2.53 & - & 1.47 & - \\
      BigBind & 60.80 & - & - & 3.82 & - \\
      DrugCLIP & 57.17 & 6.23 & 8.56 & 5.51 & 2.27 \\
      \midrule
      S-MolSearch\textsubscript{{0.4}} & 57.34 & \underline{7.58} & \underline{10.93} & \underline{6.28} & \underline{2.47} \\
      S-MolSearch\textsubscript{{0.9}} & \textbf{61.78} & \textbf{8.48} & \textbf{11.97}& \textbf{7.36} & \textbf{3.21} \\
      \bottomrule
    \end{tabular}
\end{table}

Tables~\ref{dude} and ~\ref{pcba} respectively present the performance of S-MolSearch on DUD-E and LIT-PCBA compared with other competitive baselines, where the best results are highlighted in bold and the second-best results are underlined.
The methods in the upper part of the two tables are ligand-based virtual screening methods, and their results come from~\cite{jiang2021comprehensive}. The methods in the middle part of the two tables are structure-based virtual screening methods, and their results come from DrugClip. We also present the results of S-MolSearch trained on data filtered with 0.4 and 0.9 similarity thresholds. Following previous work, we choose AUROC, BEDROC, Enrichment factor (EF) as performance metrics to evaluate both general accuracy and screening capacity, with higher values indicating better performance. Their definitions are in appendix~\ref{app:metrics}. The zero-shot setting means inferring directly without using any data from the benchmarks for training, which more close to real virtual screening scenarios.

Table~\ref{dude} shows that S-MolSearch achieves the best on all metrics. S-MolSearch trained with a strict 0.4 similarity threshold avoid overfitting similar targets and surpasses all ligand-based and structure-based virtual screening baselines. 
S-MolSearch trained with a 0.9 similarity threshold shows substantial improvements over existing methods, with over a 49\% increase in BEDROC and more than a 30\% boost in EF compared to the best baseline.
We find that S-MolSearch, as a ligand-based virtual screening method, demonstrates strong performance without requiring specific protein structures.

S-MolSearch also achieves SOTA on LIT-PCBA as shown in Table~\ref{pcba}. While its AUROC performance is not the best at the 0.4 similarity threshold, S-MolSearch perform better in BEDROC and EF, indicating its strength in screening scenarios. We notice that the metrics for all methods decline on LIT-PCBA compared to DUD-E. Unlike DUD-E, which uses putative decoys, LIT-PCBA is based on experimental results. Since many of its assays are cell-based rather than target-specific, there is noise in the active molecules, which we consider may lead to the decline. 
Meanwhile, S-MolSearch demonstrates its advantage over structure-based virtual screening by not requiring specific target information, but instead performing searches based on active molecules.

\subsubsection{Ablation Study}

\begin{table}[ht]
\centering
\caption{Ablation studies performance on DUD-E and LIT-PCBA.}
 \label{abla}
\begin{tabular}{ccc|lll|lll}
\toprule
   \multirow{2}{*}{Soft label} &  \multirow{2}{*}{Regularizer} & \multirow{2}{*}{Pretrain}   &  \multicolumn{3}{c}{DUD-E}    &  \multicolumn{3}{c}{LIT-PCBA} \\
 & & & EF 0.5\% & EF 1\% &EF 5\% & EF 0.5\% & EF 1\% &EF 5\%\\
\midrule
\ding{55} & \ding{51} & \ding{51}   & 37.35	 & 30.73  &  10.43  & 10.59 & 6.19  & \textbf{2.72}\\
\ding{51}  & \ding{55}  & \ding{51}  & 39.64 & 33.32  &  \textbf{11.47}  &   9.01 & 5.24 & 2.38   \\
\ding{51}  & \ding{51}  & \ding{55}  & 38.09 & 	31.86 & 10.88     & 8.26 &  5.24 &  2.30\\
\ding{51}  & \ding{51} & \ding{51}   & \textbf{40.85} & \textbf{34.60} & \underline{11.44} &  \textbf{10.93}  & \textbf{6.28}   & \underline{2.47}   \\
\bottomrule
\end{tabular}
\end{table}

We conduct extensive ablation studies to explore how S-MolSearch works. These results are derived from S-MolSearch trained with a 0.4 similarity threshold. First, we performed ablation studies on several key techniques of S-MolSearch. The results are summarized in Table~\ref{abla}, where the best results are bolded and the second-best results are underlined. 'Soft label' refers to training the encoder \(g_{\psi}\) using soft labels obtained from inverse optimal transport. Without this, we directly use the similarity matrix as the hard label for training. 'Regularizer' indicates the use of KoLeo regularizer. 'Pretrain' refers to starting the training of S-MolSearch from a pretrained checkpoint of Uni-Mol. Otherwise, it starts from random initialization. The results show that each component contributes to the final results. Although S-MolSearch may not be the best in some individual metrics, the absence of these components leads to poor performance on at least one benchmark. For example, not using Soft label significantly degrades performance on DUD-E. S-MolSearch performs consistently on both DUD-E and LIT-PCBA, with nearly all metrics being the best. 

\begin{figure}[ht]
    \centering
    \begin{subfigure}{0.5\textwidth}
        \centering
        \includegraphics[width=\linewidth]{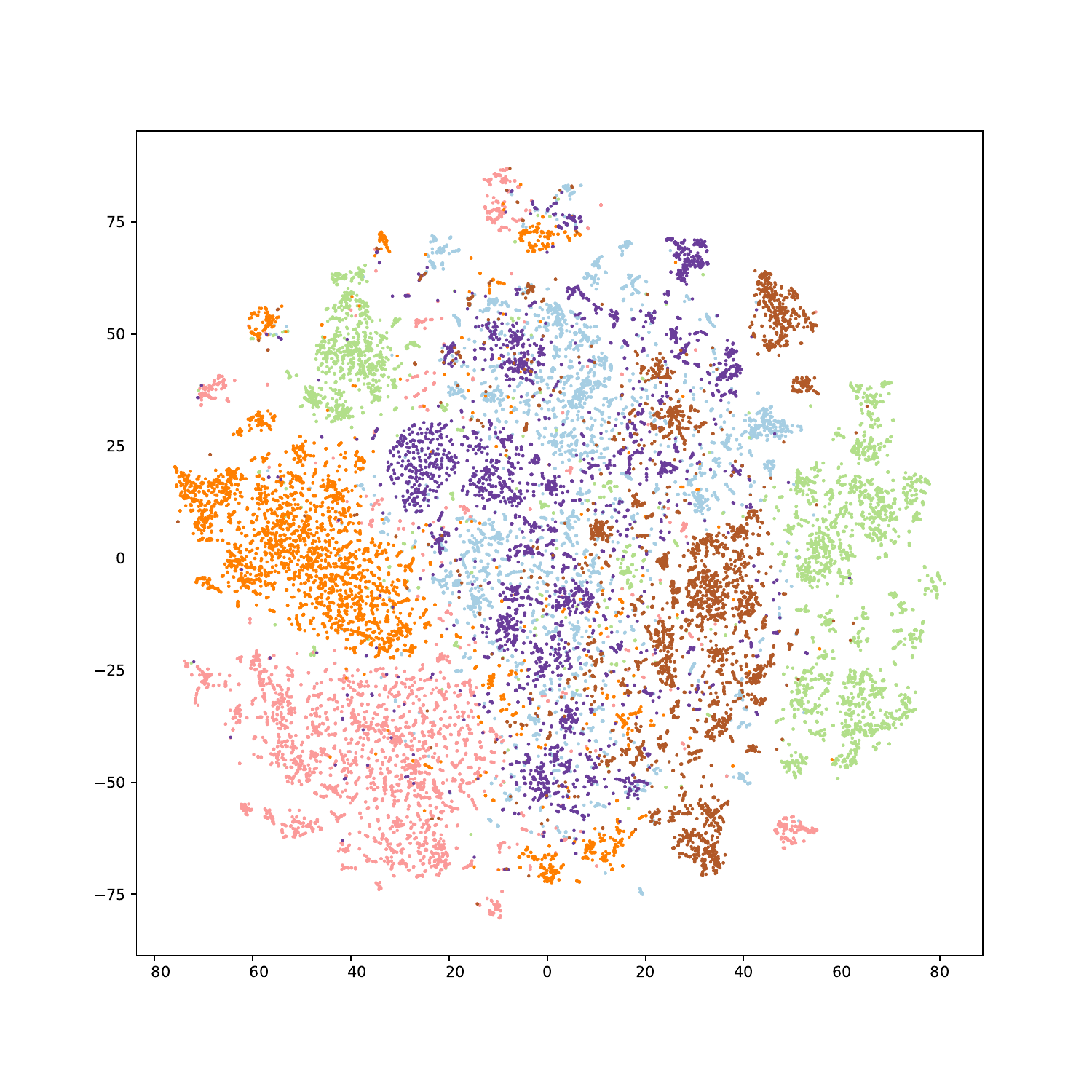} 
        \caption{Representations from pretrained checkpoint}
        \label{fig:pretrained_repr}
    \end{subfigure}%
    \begin{subfigure}{0.5\textwidth}
        \centering
        \includegraphics[width=\linewidth]{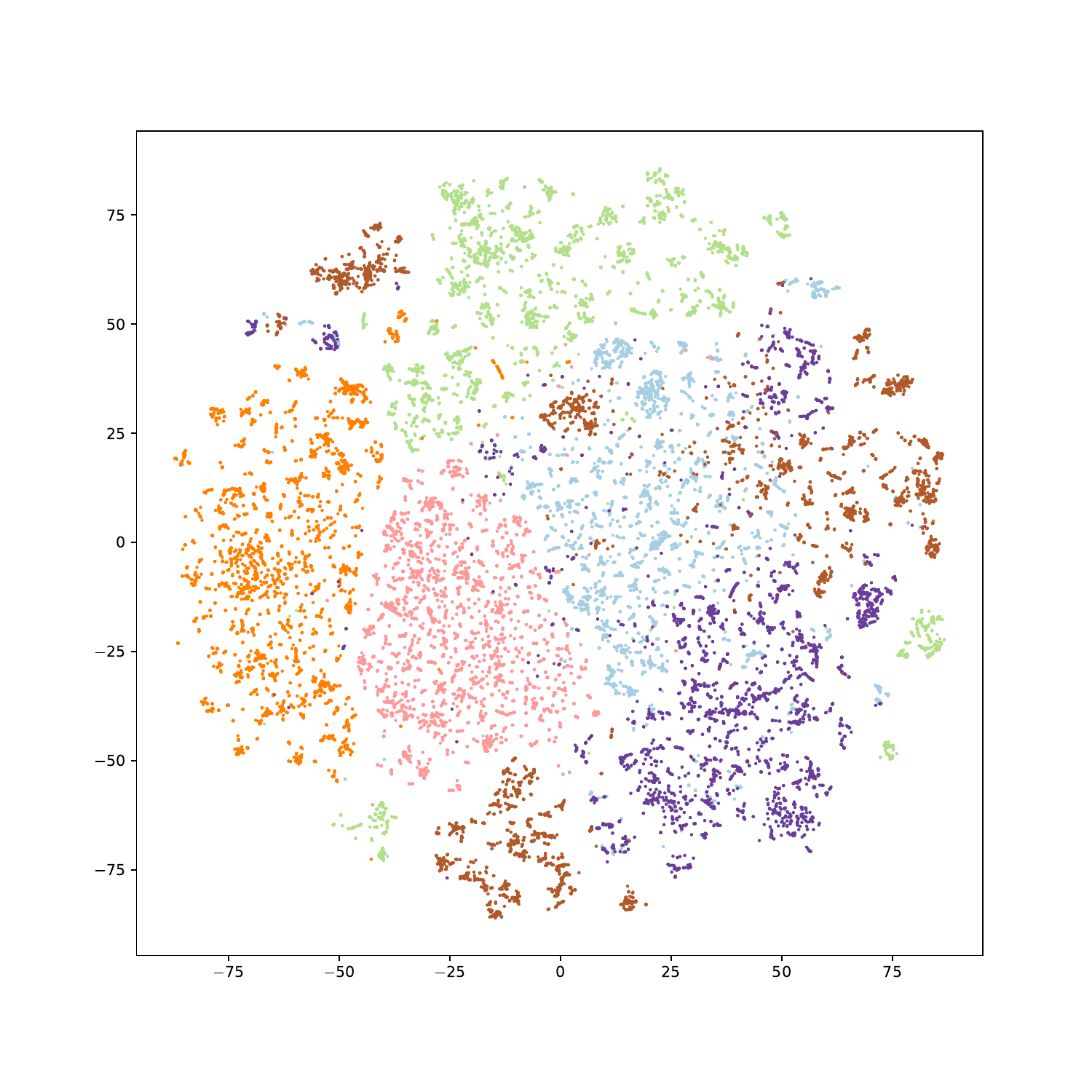}
        \caption{Representations learned by S-MolSearch}
        \label{fig:learned_repr}
    \end{subfigure}

    \caption{t-SNE visualization of molecular representations learned by S-MolSearch versus pretrained checkpoint. Different colors represent different protein targets' active molecules.}
    \label{fig:tsne_repr}
\end{figure}

To visually illustrate the difference between the embeddings learned by S-MolSearch and those from Uni-Mol, we visualize their embeddings, as shown in Figure~\ref{fig:tsne_repr}. The molecules are from ChEMBL, with different colors indicating different protein targets. Comparing the two, the classification boundaries in Figure~\ref{fig:learned_repr} from S-MolSearch are clearer, and the intra-class molecular distances are more appropriate. Some clusters split into several subclusters, possibly reflecting the hierarchical structure within the molecules.

\begin{table}[th]
  \caption{Performance under different learning paradigms on DUD-E and LIT-PCBA.}
  \label{learn_para}
  \centering
   \small
   \begin{adjustbox}{max width=\linewidth}
  \begin{tabular}{l|lll|lll}
    \toprule
    & \multicolumn{3}{c}{DUD-E}    &  \multicolumn{3}{c}{LIT-PCBA}   \\
    & EF 0.5\%  &  EF 1\%  &  EF 5\%  & EF 0.5\%  &  EF 1\%  &  EF 5\%\\

    \midrule
    Self-supervised &27.33		&20.13	& 6.81	&  4.78	&3.26   &  1.97 \\
    Supervised  &34.61		& 28.51	& 9.83	  &  7.55  &4.73   &  1.98 \\
    Finetuning &35.11 & 29.20	    & 9.96	   & 7.01	& 5.71	&2.44   \\
    \midrule
    S-MolSearch & \textbf{40.85}  & \textbf{34.60} & \textbf{11.44}   &  \textbf{10.93}  & \textbf{6.28}   & \textbf{2.47}   \\
    \bottomrule
  \end{tabular}
  \end{adjustbox}
\end{table}

In addition, we conduct ablation studies on the semi-supervised learning paradigm of S-MolSearch. The results are summarized in Table ~\ref{learn_para}. 
For 'Self-supervised', we train the model using a self-supervised learning paradigm. Specifically, we cluster the unlabeled molecule data based on their scaffolds. Molecules from the same cluster are considered as positive pairs, while those from different clusters are considered as negative pairs, and contrastive learning is performed using the InfoNCE loss. 
For 'Supervised', we use only the ChEMBL data for supervised contrastive learning, where molecules binding to the same target are treated as positive pairs and those binding to different targets as negative pairs.
For 'Finetuning', we first train the model under the self-supervised paradigm described above, then, starting from the self-supervised checkpoint, perform the supervised learning described above on the ChEMBL data.
The results show that S-MolSearch consistently performs well on both benchmarks, achieving the best results in almost all metrics. Notably, compared to finetuning, S-MolSearch demonstrates a superior ability to integrate information from both unlabeled and labeled data in ligand-based virtual screening scenarios.

\subsubsection{Impact of Labeled Data Scale}

\begin{figure}[ht]
    \centering
    \begin{subfigure}{0.5\textwidth}
        \centering
        \includegraphics[width=\linewidth]{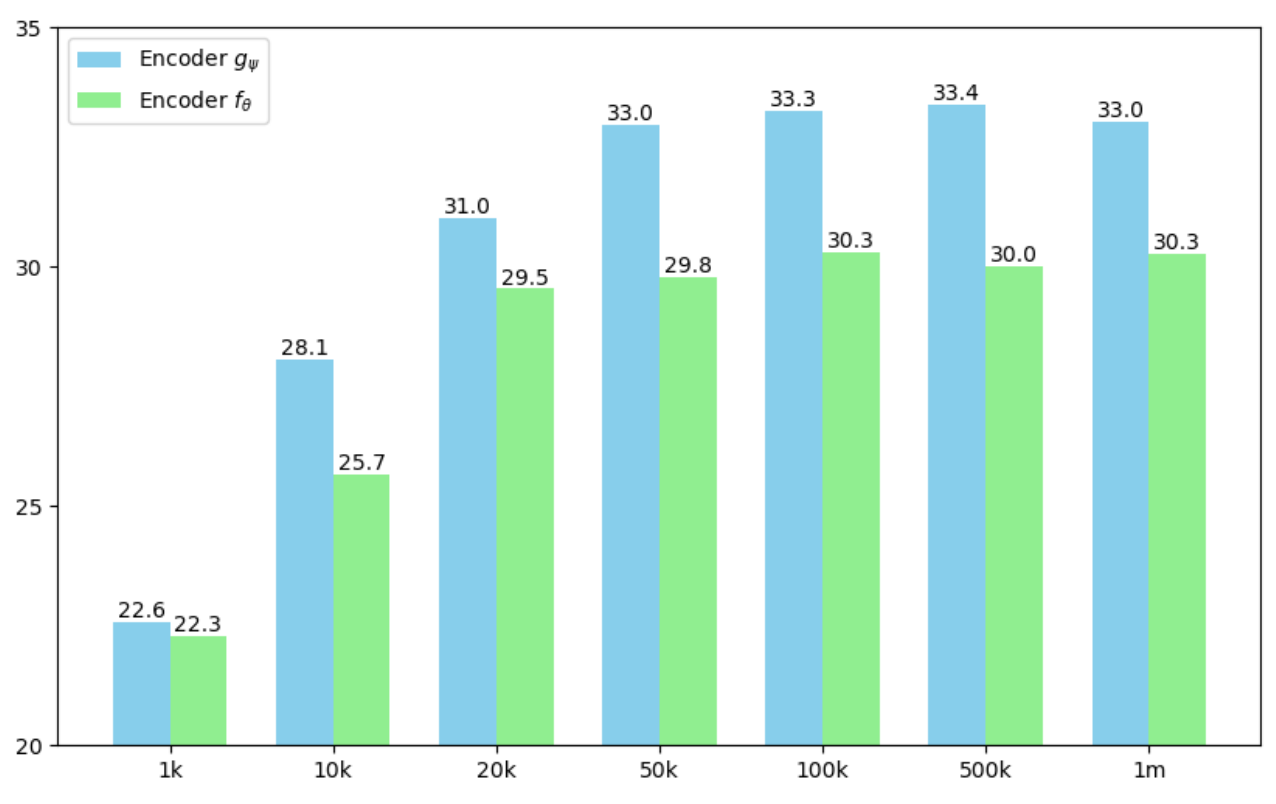} 
        \caption{EF 1\% on DUD-E}
        \label{fig:scale_dude}
    \end{subfigure}%
    \begin{subfigure}{0.5\textwidth}
        \centering
        \includegraphics[width=\linewidth]{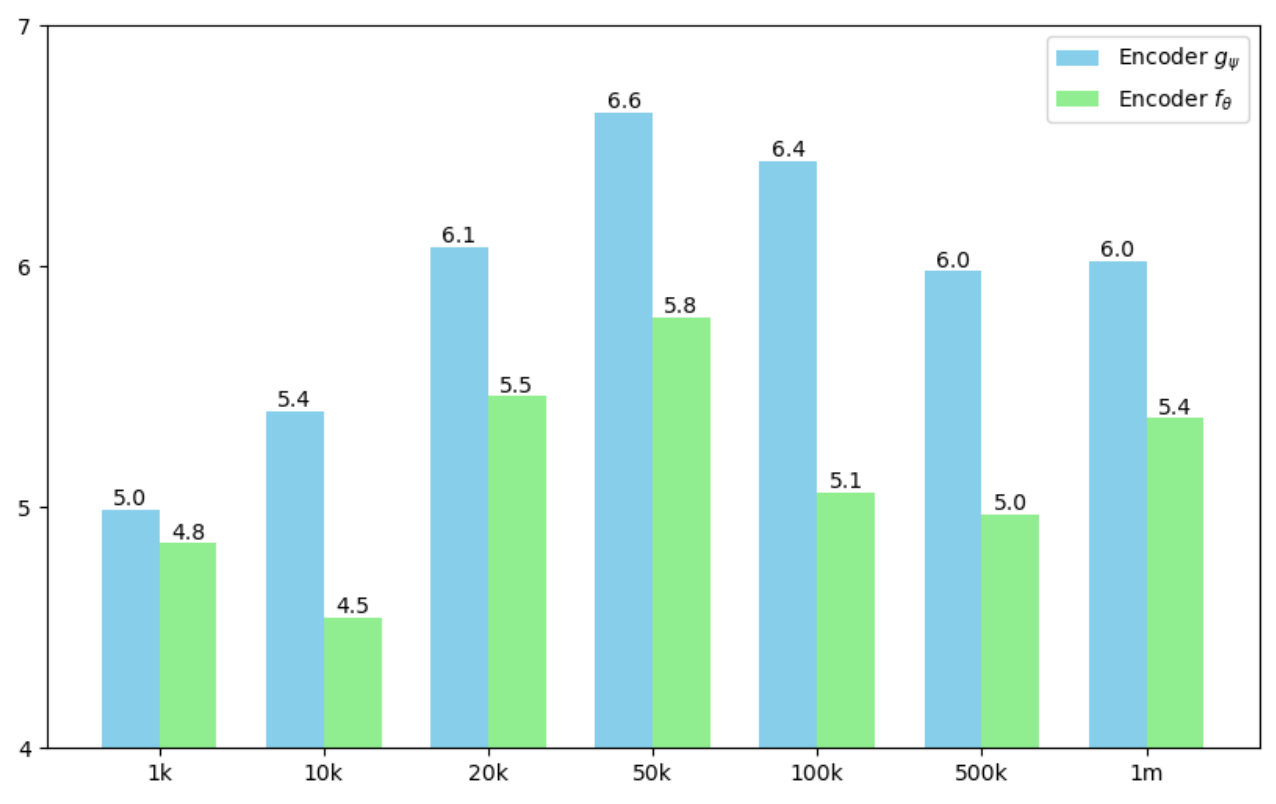} 
        \caption{EF 1\% on LIT-PCBA}
        \label{fig:scale_pcba}
    \end{subfigure}
    \caption{Performance on DUD-E and LIT-PCBA with varying numbers of labeled data, while keeping unlabeled data fixed at 1m. The blue bars represent the results of encoder \(g_{\psi}\), while the green bars represent the results of encoder \(f_{\theta}\).}
    \label{fig:scale}
\end{figure}

In the molecular field, obtaining or creating labeled data can be expensive. We also analyze how the scale of labeled data affects the results. These results are derived from S-MolSearch trained with a 0.4 similarity threshold. As shown in Figure~\ref{fig:scale}, experiments are conducted with varying amounts of labeled data, while keeping the unlabeled data fixed at 1 million. 
The performance of encoder \(g_{\psi}\) improves as the amount of labeled data increases, especially when the absolute number of labeled data is limited. Additionally, the results of encoder \(g_{\psi}\) are consistently higher than encoder \(f_{\theta}\) trained using only labeled data. 
The best results are achieved with 50k and 100k labeled data, corresponding to labeled-to-unlabeled data ratios of 1:20 and 1:10, respectively. Beyond these amounts, increasing labeled data results in stable or slightly declining performance, suggesting that further improvements may necessitate an increase in unlabeled data or a reevaluation of hyperparameters. This trend helps us to find an optimal balance between labeled and unlabeled data to maximize efficiency and performance in S-MolSearch. 

\section{Conclusion}

This study introduces S-MolSearch, a novel semi-supervised contrastive learning framework that significantly enhances the generalizability of machine learning models in virtual screening. Built on inverse optimal transport, S-MolSearch skillfully integrates limited labeled data with a vast reservoir of unlabeled data and excels at identifying potential drug candidates from extensive molecular libraries, substantially improving the accuracy and efficiency of molecule searches.
This advancement addresses current challenges in virtual screening by facilitating efficient filtering of large datasets, highlighting the framework's capability in scenarios where data annotation is costly.

Currently, S-MolSearch predominantly focuses on the molecular affinity data, omitting broader biochemical interactions, which suggests a potential area for improvement. Future work could integrate more extensive unsupervised datasets to further refine the framework's effectiveness and explore additional applications in various bioinformatics fields.

\begin{ack}

This research was supported in part by National Natural Science Foundation of China (No. U2241212, No. 61932001), by National Science and Technology Major Project (2022ZD0114802), by Beijing Natural Science Foundation (No. 4222028), by Beijing Outstanding Young Scientist Program No.BJJWZYJH012019100020098. This research was also an outcome of “AI-Aided Drug Design Based on Universal Representation of Multi-Modal Graph Structures" (RUC24QSDL014), funded by the "Qiushi Academic - Dongliang" Talent Cultivation Project at Renmin University of China in 2024. We also wish to acknowledge the support provided by the fund for building world-class universities (disciplines) of Renmin University of China, by Engineering Research Center of Next-Generation Intelligent Search and Recommendation, Ministry of Education, Intelligent Social Governance Interdisciplinary Platform, Major Innovation \& Planning Interdisciplinary Platform for the “Double-First Class” Initiative, Public Policy and Decision-making Research Lab, and Public Computing Cloud, Renmin University of China. The work was partially done at Gaoling School of Artificial Intelligence, Beijing Key Laboratory of Big Data Management and Analysis Methods, MOE Key Lab of Data Engineering and Knowledge Engineering, and Pazhou Laboratory (Huangpu), Guangzhou, Guangdong 510555, China.

\end{ack}

{
\small
\bibliographystyle{unsrt}  
\bibliography{reference}

}

\newpage
\appendix

\begin{appendix}

\section{Data Curation Details}
\label{app:chembl_data}
The data we used are manually extracted from regularly published primary literature, and then further curated and standardized. 
The whole ChEMBL database of version 33 is downloaded and cleaned via the following steps. The data is selected with assay type of 'B', target type of 'SINGLE PROTEIN', molecule type of 'Small molecule', standard type of Ki, Kd, IC50, EC50 in nM, and standard relation of '=' and '<'. Moreover, data with abnormal concentration value is further deleted, such as negative and unreasonable large ones.

\section{Metrics}
\label{app:metrics}

The BEDROC metric is designed to assess early retrieval performance, giving higher weights to active compounds that are ranked closer to the top. It is defined as:

\[
\text{BEDROC}_{\alpha} = \frac{\sum_{i=1}^{n} e^{-\alpha r_i / N}}{R_\alpha \left( \frac{1 - e^{-\alpha}}{e^{\alpha / N} - 1} \right)} \times \frac{R_\alpha \sinh(\alpha / 2)}{\cosh(\alpha / 2) - \cosh(\alpha / 2 - \alpha R_\alpha)} + \frac{1}{1 - e^{\alpha (1 - R_\alpha)}}
\]

where \( n \) is the total number of active compounds, \( N \) is the total number of molecules, \( r_i \) is the rank of the \( i \)-th active compound. Following previous work, we set \( \alpha = 85 \) to prioritize early retrieval.

We also use the Enrichment Factor (EF) to evaluate the effectiveness of virtual screening methods. The calculation formula for EF is as follows:

\[ \text{EF} = \frac{n_a/N_{x\%}}{n/N} \]

where \( n \) represents the total number of active compounds in the database, \( N \) represents the total number of molecules, \( N_{x\%} \) represents the top \(x\% \)of all molecules, and \( n_a \) represents the number of active compounds within the top \(x\% \) of molecules.

\section{Proof for S-Molsearch Induced by Inverse Optimal Transport}
\label{app:s-molsearch framework}
First, we introduce a lemma to establish the relationship between IOT and contrastive learning from \cite{shi2023understanding}.
\begin{lemma}
The optimization problem \ref{IOT-CL-lemma} and \ref{infoNCE} are equivalent:
\label{IOT-CL-lemma}
\begin{equation}
\begin{aligned}
& \min_{\theta}  KL(\hat{P}||P^{\theta})\\
& \text{subject to} \quad 
\begin{aligned}[t]
   &  P^{\theta} = \arg\min_{P \in U(a)} \langle C^{\theta}, P \rangle - \tau H(P)
\end{aligned}
\end{aligned}
\end{equation}
 where \(C^{\theta} \in R^{M \times N}, C^{\theta}(i, j) = c - s_{ij}(\theta)\) and \(\hat{P}(i, j)=\frac{\delta_{ij}}{n} \), \(\delta\) denotes the Kronecker delta function. 

\begin{equation}
\label{infoNCE}
\min_{\theta} - \sum_{i=1}^{n} log(\frac{exp(s_{ii}(\theta)/\tau)}{\sum_{j \neq i}exp(s_{ij}(\theta)/\tau)}) 
\end{equation}

In addition, \(P^{\theta}\) has the form as follows:
\begin{equation}
\label{p theta}
P^{\theta} = \frac{exp(s_{ii}(\theta)/\tau)}{\sum_{j \neq i}N exp(s_{ij}(\theta)/\tau)}
\end{equation}

\end{lemma}
\subsection{Proof for Proposition \ref{IOT-SCL-Framework}}
\label{SCL-F}
In the optimization problem \ref{eq framework}, since the setting of \(\mathcal{T}\) does not involve parameter optimization of \(\theta\) or \(\psi\), it is evident that the parameter optimization processes for \(\theta\) and \(\psi\) are independent.
According to lemma \ref{IOT-CL-lemma}, we can transform the original optimization problem into the following problem:
\begin{equation}
\label{eq framework induce}
\begin{aligned}
& \min_{\theta, \psi} \left( KL(\Gamma^{g} || \Gamma^{\theta}) + KL(\hat{\Gamma}^{\theta} || \Gamma^{\psi}) + \mu Reg_{1}(\Gamma^{\psi}) + \nu Reg_{2}(\Gamma^{\theta})) \right) \\
& \text{subject to} \quad 
\begin{aligned}[t]
    & \Gamma^{\theta} = \frac{exp(s_{ii}(\theta)/\tau)}{\sum_{j \neq i}N exp(s_{ij}(\theta)/\tau)}, \\
    & \Gamma^{\psi} = \frac{exp(s_{ii}(\psi)/\tau)}{\sum_{j \neq i}N exp(s_{ij}(\psi)/\tau)},  \\
    & \hat{\Gamma}^{\theta} = \mathcal{T}(f_{\theta}^{fixed}, g_{\psi}^{fixed}, X_{sup}, X_{full})
\end{aligned}
\end{aligned}
\end{equation}
Due to
\begin{equation}
\label{eq KL}
KL(p||q) = H(p,q) - H(p)
\end{equation}
where \(H(p)\) represents the entropy of \(p\), and \(H(p,q)\) represents the cross-entropy between \(p\) and \(q\).
Given the selection of a specific \(\mathcal{T}\) to derive \(\hat{\Gamma}^{\theta}\), we can thus determine \(\hat{\Gamma}^{\theta}(i,j)\) as a fixed quantity, making \(H(p)\) constant.
Furthermore, we can obtain
\begin{equation}
\label{eq framework induce last}
\begin{aligned}
    &\min_{\theta, \psi} \Bigg( 
        - \sum_{i=1}^{n} \log\left( \frac{\exp(s_{ii}(\theta)/\tau)}{\sum_{j \neq i} \exp(s_{ij}(\theta)/\tau)} \right) \\
        & \qquad - \sum_{i=1}^{n} \sum_{j=1}^{n} \hat{\Gamma}^{\theta}(i, j) \log\left( \frac{\exp(s_{ij}(\psi)/\tau)}{\sum_{k \neq i} \exp(s_{ik}(\psi)/\tau)} \right) \\
        & \qquad + \mu \, \mathrm{Reg}_{1}(\Gamma^{\psi}) + \nu \, \mathrm{Reg}_{2}(\Gamma^{\theta}) 
    \Bigg) \\
    &\text{where} \quad 
    \begin{aligned}[t]
        \hat{\Gamma}^{\theta}(i,j) = \mathcal{T}(f_{\theta}^{\text{fixed}}, g_{\psi}^{\text{fixed}}, X_{\text{sup}}, X_{\text{full}})_{(i, j)}
    \end{aligned}
\end{aligned}
\end{equation}

\subsection{Proof for Proposition \ref{IOT-SMOL}}
\label{SMOL}

Since \(C^{\theta^{fixed}}\) does not involve parameter optimization of \(\theta\), we can still calculate the value of \(\hat{\Gamma}^{\theta}(i,j)\). According to the proof in \ref{SCL-F}, we know that 

\begin{equation}
\label{eq framework induce molsearch}
\begin{aligned}
    &\min_{\theta, \psi} \Bigg( 
        - \sum_{i=1}^{n} \log\left( \frac{\exp(s_{ii}(\theta)/\tau)}{\sum_{j \neq i} \exp(s_{ij}(\theta)/\tau)} \right) \\
        & \qquad - \sum_{i=1}^{n} \sum_{j=1}^{n} \hat{\Gamma}^{\theta}(i, j) \log\left( \frac{\exp(s_{ij}(\psi)/\tau)}{\sum_{k \neq i} \exp(s_{ik}(\psi)/\tau)} \right) \\
        & \qquad + \mu \, \mathrm{Reg}_{1}(\Gamma^{\psi})) 
    \Bigg) \\
    &\text{where} \quad 
    \begin{aligned}[t]
        \hat{\Gamma}^{\theta}(i,j) = \arg\min_{\Gamma \in U(a,b), \, a = \mathbf{1}_{N}, \, b = \mathbf{1}_{N}} \left( \langle C^{\theta^{\text{fixed}}}, \Gamma \rangle + \frac{\lambda}{2} \|\Gamma\|^{2} \right)_{(i, j)}
    \end{aligned}
\end{aligned}
\end{equation}

When we choose $ \mathrm{Reg}_{1} $ as the regularization term~\ref{reg smol}, we obtain the optimization formulation of S-MolSearch.

\section{Reproduce details}

For the training of S-MolSearch , we use Adam optimizer at a learning rate of 0.001. The batch size is 128, and the training is conducted on 4 NVIDIA V100 32G GPUs. For backbone model, we use the same parameters as Uni-Mol. To enhance the training, we retain the masking and coordinate noise addition for atoms within molecules, as implemented in Uni-Mol. The parameters are identical to those in Uni-Mol, with a masking ratio of 0.15 and noise following a uniform distribution between -1 and 1 \AA. We randomly sample data from the labeled and unlabeled datasets to form the validation set, and select checkpoints based on the loss. More detailed configurations can be found in the code repository.

\section{Extending to the Few-Shot Setting}

Extending S-MolSearch from a zero-shot to a few-shot setting is feasible. Due to the lack of a universal few-shot setup standard in this scenario, we explore two few-shot settings. For data splitting in both settings, we randomly selected 70\% of the active molecules from each target in DUD-E as the training set for few-shot learning, while the remaining 30\% and all inactive molecules serve as test data. The initial training set includes approximately 16,000 molecules, while the test dataset contains around 1,418,000 molecules. From these 16,000 active molecules, we randomly selected 50,000 pairs as the contrastive learning training data. In the first setting, we simply add the query molecule to the active molecule set corresponding to the target and randomly sample pairs, denoted as R in Table \ref{app:fewshot}. Pairs involving molecules bound to the same target are treated as positive, while those involving molecules bound to different targets are treated as negative. In the second setting, we fix one side of each contrastive learning pair as the query molecule, denoted as F in Table. If the other molecule in the pair is an active molecule bound to the same target, it is considered a positive pair; if bound to a different target, it is considered a negative pair. From Table \ref{app:fewshot}, it can be seen that S-MolSearch performs better in the few-shot setting than in the zero-shot setting, indicating its potential in few-shot scenarios.

\begin{table}[ht]
  \centering
    \caption{Performance on DUD-E in two few-shot settings.}
    \small
    \label{app:fewshot}
    \centering
    \small
    \begin{tabular}{l|l|lll}
      \toprule
      Method & AUROC (\%) & EF 0.5\% & EF 1\% & EF 5\% \\
      \midrule
      $ \text{R}_{\text{zero-shot}}$ & 85.38 & 79.08  & 47.12 & 11.82 \\
      $ \text{R}_{\text{few-shot}}$ & 97.21  & 154.90 & 86.00 & 18.48 \\
      
      \midrule
      $ \text{F}_{\text{zero-shot}}$ & 84.87 & 79.07  & 46.45 & 11.70 \\
      $ \text{F}_{\text{few-shot}}$ & 98.32 & 165.09 & 89.98  & 19.07 \\
      \bottomrule
    \end{tabular}
\end{table}

\section{Qualitative Examples of Similarities}

\begin{figure}[t]
    \centering
    \includegraphics[width=\textwidth]{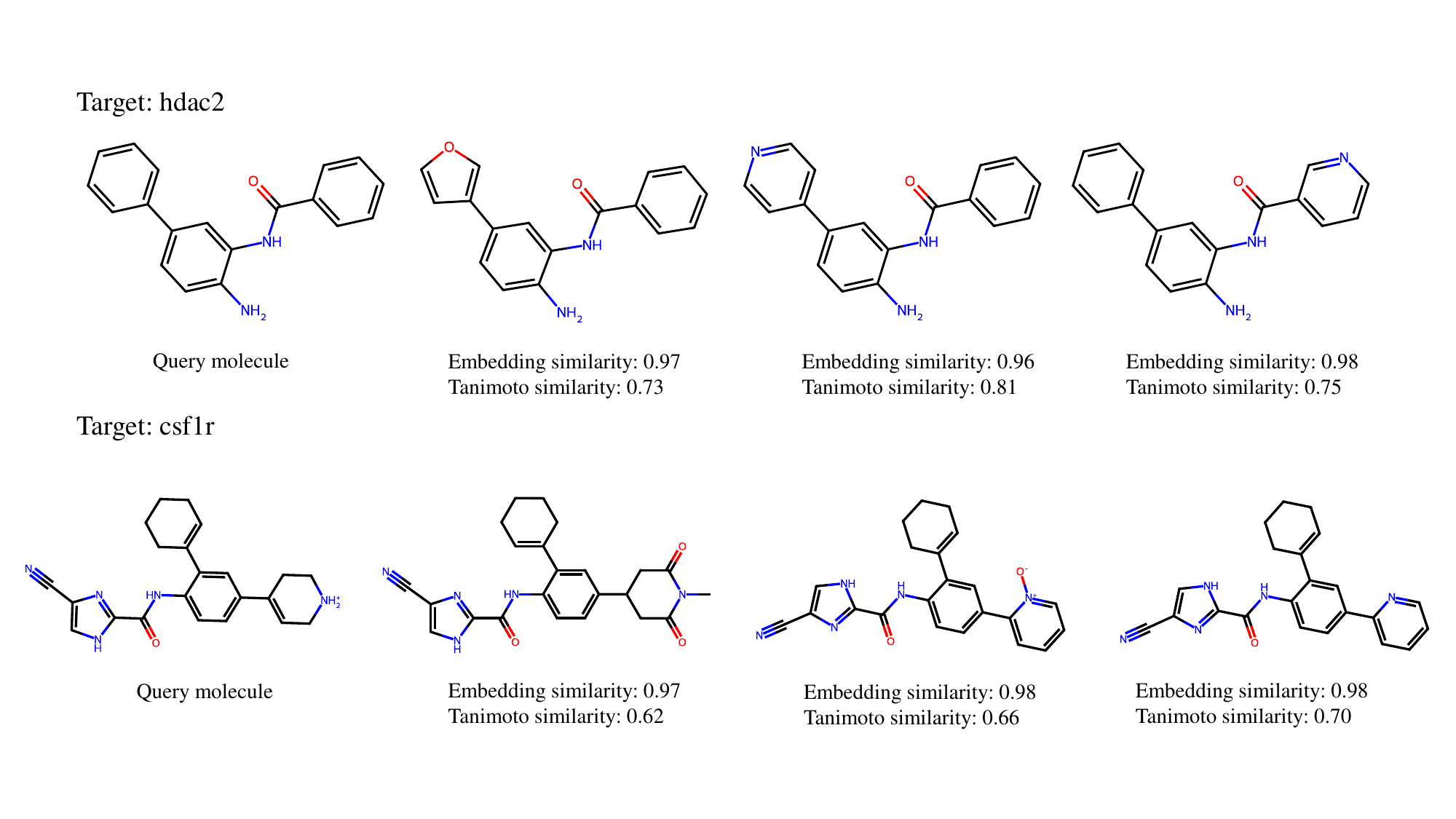}
    \caption{Qualitative examples of similarities for targets hdac2 and csf1r in DUD-E.}
    \label{fig:tani}
\end{figure}

We provide qualitative examples to enhance understanding of S-MolSearch's capabilities. Specifically, we select two targets, hdac2 and csf1r, from DUD-E. In Figure \ref{fig:tani}, we present the query molecules along with the top-ranked molecules retrieved by S-MolSearch, all of which are active molecules. We provide the embedding similarity and the Tanimoto similarity of molecular fingerprints between these molecules and the query molecule. The results indicate that molecules with high Tanimoto similarity also tend to have high embedding similarity.

\graphicspath{{figures/}}

\section{Limitations}

S-MolSearch falls short in terms of interpretability. Traditional 3D molecule search methods can capture shape and functional features required for biological interactions, providing scientists with mechanistic insights. S-MolSearch is currently unable to provide such intuitive insights for molecule search. Additionly, its use of two encoders also leads to higher memory consumption than single-encoder methods.

\section{Potential societal impacts}

S-MolSearch performs well on LBVS and can help screen bioactive molecules from large molecule databases. This eases the workload of medicinal chemists and may accelerate the discovery of new drug. On the downside, S-MolSearch also runs the risk of being used inappropriately, such as when it is used to search for similar molecules to drugs that are addictive.

\end{appendix}


\end{document}